\documentclass[runningheads]{llncs}

\usepackage{graphicx}
\usepackage{hyperref}
\usepackage{amsmath}
\usepackage{amssymb}
\usepackage{booktabs}

\usepackage{csquotes}
\usepackage{subcaption}
\usepackage{microtype}

\usepackage{tabularx}
\usepackage{array}
\newcolumntype{Z}{>{\centering\let\newline\\\arraybackslash\hspace{0pt}}X}

\def\ppcalc{\texorpdfstring{\texttt{ppcalc}\xspace}{ppcalc}}
\def\TorFS{\texorpdfstring{\texttt{TorFS}\xspace}{TorFS}}

\def\clap#1{\hbox to 0pt{\hss#1\hss}}

\usepackage{xspace}
\newcommand*{\eg}{e.g.\@\xspace}
\newcommand*{\ie}{i.e.\@\xspace}

\usepackage{extdash}

\usepackage[backend=biber,
  mincrossrefs=1000,
  style=lncs,
  doi=false,
  isbn=false,
  url=false,
  eprint=false
]{biblatex}

\AtEveryCitekey{\clearfield{pages}}
\AtEveryBibitem{\clearfield{pages}}

\AtEveryCitekey{\clearfield{location}}
\AtEveryBibitem{\clearfield{location}}

\begin{document}

\title{Progressive Pruning: Analyzing the Impact of Intersection Attacks}
\titlerunning{Progressive Pruning}
\author{Christoph Döpmann\inst{1,2} \and
Maximilian Weisenseel\inst{2} \and
Florian Tschorsch\inst{2}}
\institute{%
Technische Universität Berlin, Germany \\%
\email{christoph.doepmann@tu-berlin.de}%
\and
Technische Universität Dresden, Germany \\
\email{\{maximilian.weisenseel,florian.tschorsch\}@tu-dresden.de}%
}

\maketitle
\begin{abstract}

Stream-based communication dominates today's Internet,
posing unique challenges for anonymous communication networks~(ACNs).
Traditionally designed for independent messages,
ACNs struggle to account for the inherent vulnerabilities of streams,
such as susceptibility to intersection attacks.
In this work, we address this gap and introduce \emph{progressive pruning},
a novel methodology for quantifying
the susceptibility to intersection attacks.
Progressive pruning quantifies and monitors anonymity sets over time,
providing an assessment of an adversary's success
in correlating senders and receivers.
We leverage this methodology to analyze synthetic scenarios
and large-scale simulations of the Tor network
using our newly developed \TorFS simulator.
Our findings reveal that anonymity is significantly influenced
by stream length, user population, and stream distribution across the network.
These insights highlight critical design challenges for future ACNs
seeking to safeguard stream-based communication
against traffic analysis attacks.

\end{abstract}
\section{Introduction} \label{sec:intro}

Anonymous communication is becoming increasingly vital in today's society,
as it aims to prevent the correlation of senders and receivers in communication networks.
While early anonymous communication networks~(ACNs) focused on high-latency networks
operating with \emph{independent messages},
the dominant model today involves low-latency networks
designed to transport \emph{streams} of data.
A notable example is the Tor network~\cite{dingledine04tor}.
However, a fundamental vulnerability of such systems lies in \emph{traffic analysis} attacks.
In these attacks, adversaries monitor (encrypted) traffic
and use observed timing information to correlate and link senders with receivers.
Recent incidents, such as the German Federal Criminal Police Office~(BKA)
successfully executing a timing attack~\cite{panorama-boystown},
highlight the pressing need to address these vulnerabilities.
Dismissing such attacks as out of scope is no longer a viable option.

In this paper, we establish a deeper understanding
of how stream-based communication impacts susceptibility to traffic analysis.
For this, we introduce \emph{progressive pruning},
a methodology for quantifying anonymity sets under \emph{intersection attacks},
specifically designed for analyzing \emph{streams of communication}.
Progressive pruning works by combining anonymity sets of individual messages
as they occur over the course of a stream,
into a common \emph{pruned} anonymity set for the entire stream.
This approach enables estimating an attacker's success chances
when performing an intersection attack on a monitored stream.
For instance, progressive pruning can be used to compare the susceptibility
of different system variants to such attacks.
We argue that adopting this perspective is crucial
for designing future anonymity systems capable of providing
robust anonymity for stream-based communication.

We make use of progressive pruning to derive general, asymptotic anonymity trends from stream properties.
Our results indicate that anonymity is primarily influenced
by the relative stream length among all streams,
as well as the number of streams.
We show that long streams implicitly take the role of cover traffic for others
while experiencing poor anonymity themselves.
Applying progressive pruning to Tor as a real-world network,
we find that anonymity is distributed very unevenly.

Our contributions in this work are the following:
\begin{itemize}
\item
We develop a methodology called \emph{progressive pruning} (Section~\ref{sec:model})
to systematically evaluate the anonymity properties of streams, \ie, sequences of messages,
as observed by an adversary who monitors traffic entering and leaving an ACN.
We also discuss anonymity properties that can be derived.

\item
We apply our optimized implementation of progressive pruning (\ppcalc)
to evaluate the anonymity impact of different stream properties
based on synthetic network traces (Section~\ref{sec:synthetic-streams}).

\item
We demonstrate how progressive pruning can be applied to a large-scale practical system
by analyzing anonymity implications of the network topology in Tor.
We carry out large-scale simulations of traffic flows in today's Tor network
as well as scaled variants (Section~\ref{sec:scaling-tor}).
For this, we introduce the \emph{Tor Flow Simulator (\TorFS)},
a scalable alternative to the well-known TorPS~\cite{johnson2013usersrouted}
that adds the ability to simulate traces of traffic flows.
\end{itemize}

\section{Related Work}

The study of anonymous communication began in 1981,
when Chaum introduced the concept of a~mix~\cite{chaum1981untraceable}.
The key idea of achieving unlinkability through \emph{independent messages} of uniform size,
making them indistinguishable to observers,
has since become foundational for mixnet designs,
including modern systems like Loopix~\cite{piotrowska2017loopix}
and Nym~\cite{diaz2021nym}.
The anonymization of \emph{streams} of data gained prominence
with the introduction of \emph{onion routing}~\cite{syverson1997anonymous},
which became a core design principle in Tor~\cite{dingledine04tor}.
Other low-latency anonymity networks have since adopted
this approach~\cite{chen2015hornet,DBLP:journals/popets/ChenP17}.

The major weakness of low-latency data streams lies
in their vulnerability to traffic analysis,
particularly \emph{intersection attacks}~\cite{berthold2001disadvantages}.
Correlation techniques range from simple packet counting~\cite{DBLP:conf/ih/BackMS01}
to advanced deep learning methods~\cite{DBLP:conf/ccs/NasrBH18}.
Temporal characteristics have also been exploited for de\Hyphdash*anony\-mi\-zation
by linking multiple streams of the same user~\cite{johnson2013usersrouted}.
While effective in practice, these approaches often fail to clarify how stream properties,
such as length or concurrency, impact anonymity.
Our methodology builds on general intersection attacks
to study the influence of inherent stream properties on anonymity.

\section{Progressive Pruning} \label{sec:model}

In this section, we introduce \emph{progressive pruning},
a novel methodology designed to analyze the anonymity implications of data streams
in anonymous communication networks~(ACNs).

\subsection{Adversary Model}

\begin{figure}[b]
    \centering
    \includegraphics{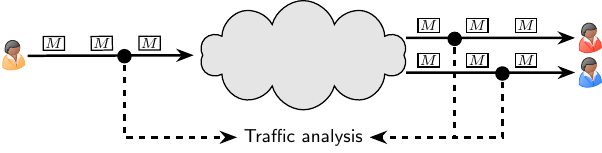}
    \caption{
      For some user~(yellow), an attacker aims to identify the communication partner~(red or blue)
        using traffic analysis.
    } \label{fig:acn-overview}
\end{figure}

In order to evaluate the anonymity properties of stream-based communication,
we assume that an attacker can monitor both ends of the streams.
That is, the attacker can observe when packets of each stream are \emph{sent} and \emph{received}.
This setting, therefore, focuses on \emph{traffic analysis} attacks
in which the attacker compares temporal traffic patterns at both ends of the streams,
trying to identify pairs of communication~(see Figure~\ref{fig:acn-overview}).
In the context of stream-based communication,
the primary attack is an \emph{intersection attack}~\cite{berthold2001disadvantages},
which will therefore be the main concern of this work.
The idea behind such attacks is to exploit the knowledge of
multiple messages belonging together by intersecting their anonymity sets
in order to identify the communication partners.
Our adversary model is inspired by a global passive adversary,
which appears to be a sensible assumption today~\cite{nsa-buys-data},
and deliberately abstracts from any concrete underlying communication system
to capture the anonymity properties inherent to streams.

\subsection{Overview}

Progressive pruning builds on the well-known notion of anonymity sets---%
but \emph{evolves over time:}
Given a network trace of messages
with observation timestamps at their source and destination,
we model the core concept of an intersection attack
and quantify anonymity sets per stream.
Initially, a message may have a large anonymity set as
there may be a broad set of candidates.
However, as communication progresses and more messages are exchanged,
the anonymity set shrinks as candidates are excluded.
For example, the observed timing of events may conflict
with other previously gathered observations,
rendering them incompatible.
Pruning the set of candidates progressively
gives our methodology its name.
We illustrate progressive pruning for a single message
in Figure~\ref{fig:window-model}.

We note that, as a consequence of the design,
our anonymity metric does not provide an absolute value of system security or anonymity,
which would be very hard or likely impossible.
However, it can be used to analyze the relative changes to anonymity
induced by different system parameters.

\subsection{Modeling Intersection Attacks} \label{sec:model-details}

\paragraph{Stream Model:}
We assume that the input stream traces stem from a communication system of the following abstract notion:
Unidirectional \emph{streams of messages} are exchanged between \emph{senders} and \emph{receivers}.
Messages can only be linked based on timing information and share a uniform size.
We further assume that there is no message loss,
and the delay of each message (the time between being sent and received)
is bounded between $[D_{min},D_{max}]$.
Depending on whether we want to evaluate the anonymity of senders or receivers,
either (1)~senders send all their messages to the same receiver,
or (2)~receivers receive all their messages from the same sender.
The attacker's goal is to link senders and receivers.
For clarity and without loss of generality,
we focus on the sender's perspective.
Consequently, the attacker aims to identify the corresponding receiver for each sender.
As our deliberately simple attacker represents a concrete attack instance,
the resulting anonymity can be interpreted as \emph{upper bounds} on the achieved anonymity.
Our assumptions for the stream model allow investigating
the \emph{inherent properties} of stream communication
but are not intended to replicate any real-world system.

\paragraph{Terminology:}
Let $S$ be the set of senders and $R$ the set of receivers.
Accordingly, let $M_S$ be the set of messages sent by all senders,
and $M_R$ the set of messages received by all receivers.
$M_S[s]$ is the set of all messages sent by sender~$s \in S$
and $M_R[r]$ is the set of all messages received by receiver $r \in R$.
By~$S[m]$, we denote the sender $s \in S$ of message~$m$.
Analogously, $R[m]$ is the receiver $r \in R$ of message~$m$.
The relationship between sent and received messages can be characterized as follows:
$ \forall m_S \in M_S : \exists m_R \in M_R : m_S = m_R $.
It is the adversary's goal to find this mapping,
\ie, determine the receiver of a sent message, breaking sender-receiver unlinkabilty.
Furthermore, let $T_S[m]$ be the time when message~$m$ was sent,
and $T_R[m]$ the time when message~$m$ was received.

\begin{figure}[t]
    \centering
    \includegraphics{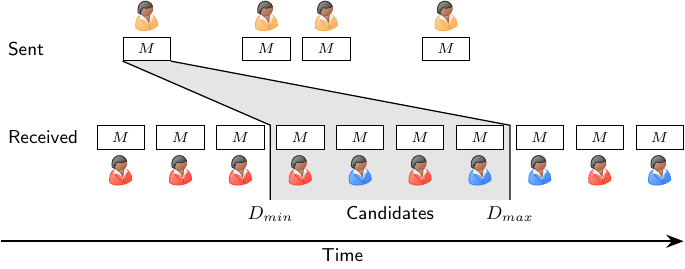}
        \caption{
            Progressive Pruning:
            For each sender message (yellow), potential receiver messages are identified
            based on the network delay bounds $D_{min}$ and $D_{max}$.
            Receivers of messages within this window (red, blue) are considered as candidates.
            The stream's anonymity set is determined by intersecting these message anonymity sets over time.
    } \label{fig:window-model}
\end{figure}

\paragraph{Anonymity set of a single message:}
We first define the anonymity set for a single message
before extending it to multiple messages afterwards (cf.~Figure~\ref{fig:window-model}).
To do so, we introduce the notion of a \emph{message-message anonymity set} $\Psi_{MM}$.
For each sent message $m_S$,
this is the set of received messages $m_R$ that $m_S$ could correspond to,
because they were received within the timeframe assumed for message delays. We therefore define it as follows:
\[
\Psi_{MM}[m_S] = 
\{ m_R \in M_R \mid T_R[m_R] \in [T_S[m_S] + D_{min}~,~T_S[m_S] + D_{max}] \}
\]
We can use this to define the \emph{message-receiver anonymity set}~$\Psi_{MR}$
for a message,
which represents the potential receiver candidates and forms a subset of~$R$:
\[
\Psi_{MR}[m_S] = \{ R[m_R]~\forall~m_R \in \Psi_{MM}[m_S] \}
\]

\paragraph{Anonymity set of multiple messages:}
We now extend these anonymity set definitions to multiple messages from the same sender---a~stream.
By doing so, we want to capture the anonymity set of senders as a whole,
taking all their messages into account.
The simplest approach would be to simply intersect the message-receiver anonymity sets
for all messages from a sender:
$ \bigcap_{m_S \in M_S[s]} \Psi_{MR}[m_S] $.
However, this falls short of the fact that the adversary
can trivially also exploit the \emph{number} of messages to differentiate between candidates.
For example, if a sender sent 10~messages in a short period of time,
but a receiver received only one~message
that falls into the message-message anonymity sets of all sent messages,
the simple intersection approach would not exclude that receiver
although it obviously cannot be a candidate because it received too few messages.

We address this issue by formulating the following assignment problem:
Let~$s$ be a sender of messages~$M_S[s]$ and $r$ a receiver of messages~$M_R[r]$.
Then, $r$ can be a candidate receiver for the messages from $s$ only
if there exists a mapping $\sigma : M_S[s] \to M_R[r]$
that assigns each sent message to a corresponding received one
and meets the following two requirements:
(1)~If $\sigma(m_S) = m_R$, then $m_R \in \Psi_{MM}[m_S]$
(messages potentially correspond w.r.t. the delay constraints), and
(2)~$\sigma$ is injective (each received message is assigned at most once).

Given the sets of sent and received messages,
it is possible to determine whether such a mapping exists,
\eg, in a brute-force manner.
In Section~\ref{sec:algorithm}, we will present an efficient algorithm
to solve this task in linear time.
If a mapping that meets the the aforementioned requirements can be found,
we say that $r$ is a \emph{candidate} for $s$.
Consequently, we can define the \emph{sender-receiver anonymity set} of sender $s$ as follows:
\[
\Psi_{SR}[s] = \{ r \in R \mid \text{$r$ is a candidate for $s$}\}
\]
This now denotes the final anonymity set usable to estimate the anonymity set of a sender.
The anonymity set size $A(s) = |\Psi_{SR}[s]|$
can consequently be regarded as a measure of the achieved anonymity.
If $A(s) = 1$, then $s$ can be de-anonymized by an adversary that observes sent and received messages.
If $A(s) > 1$, then the simple window-based attack we modeled would not be successful
because multiple candidates remain.

\subsection{Model Discussion}

The utility of progressive pruning is based on the fact that
it couples the anonymity sets of a sender's individual messages
to achieve an estimate of the sender's anonymity.
It is therefore suitable for evaluating the anonymity of stream-based communication,
as opposed to single messages.

Intuitively, we can understand the presented construction
as the anonymity set of a sender $s$ \emph{over time}.
When taking into account only a single message,
then all receivers that have received a message in the respective timeframe
are possible communication partners of $s$.
Additional messages in the stream, however, add additional constraints
that can be exploited by the adversary.
As such, more messages will eventually decrease the anonymity set.
The number of observed messages that are necessary to break anonymity
can be regarded as an indicator for the costs an adversary has to invest for the attack.

Another observation that can be made is that the number of messages sent by a sender
is highly relevant for the resulting anonymity set.
In particular, if the streams differ in their number of messages,
then the stream length is useful for an adversary to differentiate senders.
While this comes at no surprise, it has an interesting implication:
A sender that sends more data than anyone else does not retain any anonymity.
In general, the shorter the communication is in comparison to the others,
the more likely it is that a sender can \enquote{hide} among the other senders.
We also verified this important relationship formally~\cite{progressive-pruning-arxiv}.
Additionally, the absolute latency value of messages does not directly make a difference.
Instead, an attacker only benefits from a low \emph{deviation} of latency~(jitter)
because it makes the traffic more predictable, allowing use of a smaller window.

\subsection{Algorithm for Matching Candidate Messages} \label{sec:algorithm}

Here we present an efficient algorithm
that can be used to determine whether a receiver $r \in R$ is a candidate for sender $s \in S$,
based on their sets of sent and received messages, as defined in Section~\ref{sec:model-details}.
Put differently, the algorithm decides whether a suitable assignment $\sigma$ exists,
mapping the messages sent by $s$ to valid messages received by $r$.

We exploit the fact that the exact mapping is not needed,
but it is sufficient to determine if one exists.
For each sent message $m_S \in M_S[s]$, we first determine the potential candidate messages in $r$:
\[
\psi(m_S, r) =\Psi_{MM}[m_S]~\cap~M_R[r]
\]
Without loss of generality, we then iterate over the sent messages $m_S \in M_S[s]$,
ordered by their \emph{receive} timestamp.
If the messages from a sender are always delivered in order,
this corresponds to the observed send sequence.
In the special case that we evaluate a complete \enquote{white-box} network trace
containing the ground truth message mapping,
we can also determine that order.
In practice, if messages are allowed to be re-ordered,
an attacker would not know the order and might have to choose a less efficient algorithm.
This, however, is solely a matter of computational efficiency for an attacker.

During iteration, we need to decide whether $m_S$ can be mapped to a received message.
We do so by comparing $\psi(m_S, r)$ to the (previous) message anonymity set $\psi(m_S', r)$
of the previous message $m_S'$.
$m_S$ can be mapped if and only if there is at least one valid candidate message.
The number of candidates (\texttt{cand}) is determined
as the sum of leftover unused candidate messages from the previous anonymity set $\psi(m_S', r)$
that are also present in $\psi(m_S, r)$, which we call \texttt{overlap},
plus the number of new messages $|\psi(m_S, r) \setminus\psi(m_S', r)|$, which we call \texttt{new}.
If \texttt{cand > 0}, then $m_S$ can be mapped and the next message continues with \texttt{cand - 1}.
If for all messages, the number of candidates is non-zero,
then a suitable assignment $\sigma$ exists. Otherwise, this is not the case.

\subsection{Implementation} \label{sec:implementation}

We implement progressive pruning
and make it available as an open-source project called \ppcalc%
\footnote{\url{https://github.com/cdoepmann/ppcalc}}, implemented in Rust.
\ppcalc consumes network traces consisting of messages between senders and receivers, and their timestamps.
It outputs the anonymity set of each stream.
It processes plain CSV data and can thus easily be reused for other research questions.
It supports the calculation of sender anonymity as well as receiver anonymity.
During the implementation of \ppcalc, great care was taken to achieve high scalability
that enables \ppcalc to process communication traces
not only from relatively small, synthetic scenarios
but also traces from real-world networks like Tor, at scale.
For the experiments we will later present in Section~\ref{sec:scaling-tor},
\ppcalc successfully managed to process network traces of approximately 720~GB
within 74~hours on a 12~CPU~cores machine.

\section{Anonymity Factors for Streams} \label{sec:synthetic-streams}

\begin{figure}
    \begin{subfigure}{\textwidth}\centering
    \includegraphics{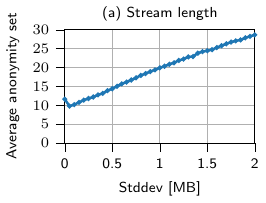}\hspace{-16pt}
    \phantomsubcaption\label{subfig:synth-length}
    \includegraphics{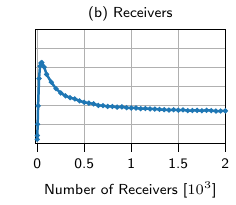}\hspace{-16pt}
    \phantomsubcaption\label{subfig:synth-topology}
    \includegraphics{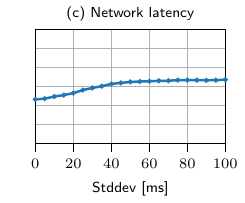}
    \phantomsubcaption\label{subfig:synth-latency}
    \end{subfigure}
    \caption{
      Influence of different stream properties on the anonymity set size.
    } \label{fig:synth-results}
\end{figure}

We now want to gain initial insight on the general factors that influence anonymity
in the case of stream communication.
By doing so, we aim to verify common assumptions and expectations
about the development of anonymity in stream communication.
Note that this investigation complements existing work
like \cite{wright2003defending} and~\cite{johnson2013usersrouted}
in that these focus on anonymity over time due to user behavior,
while we consider anonymity properties of the streams themselves.

To this end, we simulate communication between senders and receivers
with different scenario parameters
such as stream length, latency, and number of senders and receivers.
We then apply progressive pruning
and use the calculated anonymity set sizes as estimates for anonymity.
By altering the scenario parameters, we can evaluate
the relative influence of each parameter on anonymity.
Without loss of generality, we focus on sender anonymity here.
Therefore, each sender sends all of their messages to one receiver,
and the anonymity set of a sender corresponds to the set of candidate receivers as observed by an attacker.

As our base scenario, we choose 2,000~senders,
starting their transfer within a time span of 60~seconds.
This number is loosely inspired by the typical 3~million daily users in Tor.
Each sender transmits 2,314~KB of~data, corresponding to the average size of a website.%
\footnote{\url{https://almanac.httparchive.org/en/2022/page-weight\#request-bytes}}
Senders transmit at a data rate of 10\,Mbit/s, sending to a total of 400~receivers, chosen uniformly at random.
The network latency is set to 210\,ms, a typical value found in the real Tor network.%
\footnote{\url{https://metrics.torproject.org/onionperf-latencies.csv?start=2023-05-05&end=2023-08-03&server=public}}
The analysis window that bounds network latency for anonymity calculation in progressive pruning
is set to $\pm 100$~ms around this value.
For each of the parameters,
we carry out a series of simulations in which we only vary this parameter.
In the following, we describe our results,
revealing and verifying several relationships.

\paragraph{Stream Length}

Firstly, the length of a stream, in relation to other streams in the network,
is of crucial importance for the achieved anonymity (Figure~\ref{subfig:synth-length}).
The results reveal that anonymity sets increase
with growing stream length variation.
At first sight, this is counterintuitive
because more heterogeneous stream lengths may give an attacker a way to distinguish between streams.
However, what can be observed here is the effect of cover traffic:
Longer streams act as cover traffic for shorter streams,
enhancing their anonymity while sacrificing their own.
Due to the fact that a single long stream can act as cover traffic for several shorter streams,
the average anonymity is increased.
This underlines the effectiveness of cover traffic,
either implicitly through a large user base, or explicitly using designated cover traffic.

\paragraph{Network Topology}

We analyze the impact of network topology by varying the ratio between senders and receivers.
With fewer receivers, each receiver handles the streams of more senders,
leading to a higher-density network.

We find that a low number of receivers compared to the number of senders
is beneficial for anonymity because it helps streams to blend in with each other
(Figure~\ref{subfig:synth-topology}).
This is in line with our expectation
because a larger number of receivers leads to an increasingly sparse network,
in which there are fewer candidates an attacker has to differentiate.
However, a too low number of receivers is problematic for our notion of anonymity as well
because the total number of receivers constitutes an inherent upper bound for the anonymity set size.

\paragraph{Network Latency}

Lastly, we consider the case of varying network lateny~(jitter).
The results in Figure~\ref{subfig:synth-latency} indicate a small increase in anonymity
with growing network jitter.
There are two main reasons for this:
Firstly, more network jitter generally makes the network less predictable for an attacker.
And secondly, he introduction of jitter for the single messages effectively results in
the streams spanning an increased period of time during transmission.
If streams are active for a longer time, they will be active with more other streams simultaneously,
leading to an increased anonymity set.

All in all, these results come as no surprise,
but express the soundness and utility of predictive pruning to analyze anonymity properties of streams,
also verifying these relationships based on a well-defined stream model.

\section{Case Study: Comparing Tor Instances} \label{sec:scaling-tor}

So far, we demonstrated how progressive pruning
can be used for analyzing stream-based communication
and identifying factors relevant to anonymity.
The considered scenarios, however, were primarily synthetic examples
to verify our methodology and observe trends.
Now, we apply this approach to the widely used Tor network,
providing a practical example of how progressive pruning
can assess the anonymity properties of large-scale systems.

As discussed, progressive pruning quantifies anonymity relatively.
Instead of defining \enquote{Tor's anonymity level,}
we compare network instances to identify deviations in anonymity,
focusing on the implications of network growth.

The Tor network has grown significantly over time
and it is expected to continue this trend.
This growth, however, presents challenges for future network designs.
Previous work~\cite{DBLP:conf/IEEEares/DopmannT23}
analyzed \emph{scaled} Tor instances,
revealing that purely horizontal growth may harm \emph{performance}.
Here, we leverage progressive pruning
to explore the impact of scaling on network \emph{anonymity},
which is necessary as adversaries become stronger.

\subsection{Methodology}

Our general approach is to quantify anonymity sets based on message traces
obtained from network simulations.
Specifically, we simulate the Tor network to obtain a set of circuits,
along with traces of exchanged messages~(cells) and their timestamps.
We then compute anonymity sets over time using progressive pruning.
This process is first applied to the original~(unscaled) Tor network.
In addition, we generate two synthetic but representative Tor networks
scaled either vertically or horizontally by +100\%
using torsynth~\cite{DBLP:conf/IEEEares/DopmannT23}.
We conduct network simulations on these scaled networks
and apply progressive pruning to the resulting traces
to quantify anonymity sets as well.

In order to avoid sampling artifacts,
all experiments are conducted \emph{at scale},
which presents a nontrivial engineering challenge
for the scalability of the involved software components.
We base our experiments on the network consensus from 2023-10-01 at~00:00,
and simulate one hour of full network activity.

\paragraph{Tor Flow Simulation:}
Given a specific snapshot of the Tor network~(consensus),
the first step is to generate a trace of Tor network activity.
In particular, this involves generating circuits
and using them to transport data cells,
simulating network communication.

Simulating the Tor network involves a trade-off
between computational effort and realism.
The most realistic approach uses the \emph{shadow} simulator~\cite{jansen12shadow},
capable of running the unmodified Tor software.
However, running such simulations at scale 1:1 is infeasible.
While shadow enables fine-grained simulation
to investigate performance and other networking effects,
we focus on the distribution of circuits and cells across the network,
because these factors impact anonymity as considered by progressive pruning.

We therefore adopt a more abstract simulation approach,
similar to the Tor path simulator, \emph{TorPS}~\cite{johnson2013usersrouted}.
TorPS simulates circuit generation over time
and has been used to study anonymity changes before.
However, TorPS suffers from severe scalability issues.
Furthermore, it does not simulate user behavior at the packet level
and thus cannot be used to generate the network traces we require.

We developed the \emph{Tor Flow Simulator (\TorFS)},
an open-source simulator\footnote{\url{https://github.com/cdoepmann/torfs}}
designed to simulate Tor user behavior at the packet level.
\TorFS inherits much of TorPS's functionality.
However, it is implemented in Rust with a focus on scalability
and adds the missing pieces, including the modeled circuit selection rules
and users' circuit management.
\TorFS also supersedes TorPS's simplistic user model,
replacing it with the state-of-the-art model~\cite{DBLP:conf/ccs/JansenTH18}.
That is, user behavior is now modeled as a probabilistic process
(an exponential distribution) with parameters derived
from real-world measurements on the Tor network.
Moreover, while TorPS simulates only the generation and management of circuits,
\TorFS also simulates the circuits themselves
to generate network traces of the individual messages~(cells).
For this, it again makes use of the probabilistic model from~\cite{DBLP:conf/ccs/JansenTH18}.
Consequently, \TorFS uses the \emph{stream model}
as well as the \emph{packet model} from~\cite{DBLP:conf/ccs/JansenTH18},
which mimic real traffic behavior as observed on the real Tor network.
As a result, \TorFS outputs message traces ready for analysis with \ppcalc.
Please note, however, that \TorFS does not constitute a full-fledged network simulator like shadow or~ns-3.
It abstracts from networking effects in that cells entering the network
are output after a constant latency at their destination.

\paragraph{Applying Progressive Pruning to Tor:}
In our evaluation, we focus on traffic
that is directed from the exit relay to the client
as this constitutes the majority of Tor traffic.
We apply the following mapping from simulated Tor behavior
to the message model from Section~\ref{sec:model}:
Each cell that occurs in the simulation corresponds to a message in the model.
Multiple cells that form a data stream in the simulation
also constitute a single stream of messages in the sense of our model.
The sender of these messages is the according simulated exit.
For each stream, we define a receiver as the entity that receives all of this stream's messages.
This way, the stream-based anonymity can be evaluated independently for each stream in the network.

After generating message traces of Tor,
we analyze these using progressive pruning in order to calculate anonymity sets.
For this task, we again use \ppcalc.
As laid out before, anonymity sets can be calculated in one of two possible \enquote{directions}.
In our current experiment, we have a fixed (relatively small) set of exit relays
that are sending data to many clients.
Due to this 1:n relationship, it makes sense to calculate anonymity from the receiver perspective.
That is, for each receiver (client),
we calculate the set of potential senders (exits) from which their data may originate.
Analyzing anonymity in the reverse direction (the set of potential receivers per sender)
is not useful in this scenario due to the fact that exits are sending to many clients at the same time.

\subsection{Results and Discussion}

\begin{table}[t]
\scriptsize
\caption{Anonymity results from simulations of an original Tor consensus,
Tor scaled vertically by +100\%,
and Tor scaled horizontally by +100\%.}
\label{tab:case-study-metrics}
\begin{tabularx}{\linewidth}{Xccccc}
\toprule
& Original & \qquad\qquad & Vertical & \qquad\qquad & Horizontal \\
\midrule
\textbf{Fully deanonymized streams} & 795/26,407,264 & & 41/52,808,426 & & 1096/52,773,747 \\
\textbf{Avg/median anonymity set size} & 1,241.4 / 1,440 & & 1,610.8 / 1,784 & & 2,481.9 / 2,882 \\
\textbf{Stddev of anonymity set size} & 594.4 & & 416.5 & & 1,186.0 \\
\bottomrule
\end{tabularx}
\end{table}

\begin{figure}[t!]
  \begin{minipage}[t]{.32\textwidth}
    \centering
    \includegraphics{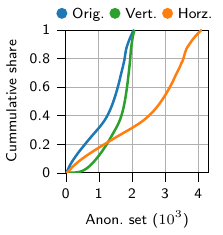}
    \caption{Anonymity sets.}
    \label{fig:case-study-cdf}
  \end{minipage}
  \hfill
  \begin{minipage}[t]{.64\textwidth}
    \begin{subfigure}{\textwidth}\centering
        \includegraphics{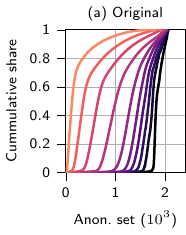}\hspace{-20pt}
        \includegraphics{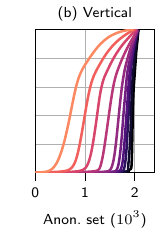}\hspace{-20pt}
        \includegraphics{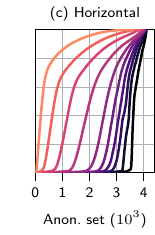}
    \end{subfigure}
  \caption{Anonymity sets by stream length (percentiles).}
  \label{fig:cdf-by-size}
  \phantomsubcaption\label{subfig:orig}
  \phantomsubcaption\label{subfig:vert}
  \phantomsubcaption\label{subfig:horz}
  \end{minipage}
\end{figure}

Our analysis is based on anonymity set sizes
and their distribution across the network.
Table~\ref{tab:case-study-metrics} displays basic statistics about the anonymity in each simulated network.
As a first observation, we can see that the anonymity set sizes are on average relatively large.
In the case of the original Tor network, \eg,
streams still remain with an average of more than 1,200~sender candidates
(out of approximately 2,400~exits).
It is noteworthy that, with regard to the simple window-based attack progressive pruning mimics,
anonymity of the vast majority of streams is not \emph{trivially} broken.
This observation is supported by the fact that---looking again at the original Tor network---%
out of 26 million streams, less than~1,000 are fully de-anonymized,
\ie, resulting in an anonymity set size of~1.

In order to better understand how anonymity is distributed within the network,
we consider CDF plots of the anonymity set sizes, displayed in Figure~\ref{fig:case-study-cdf}.
Looking at the data from the original Tor network first,
it becomes immediately apparent that anonymity is distributed very unevenly between the streams.
Note that, if all streams had the same anonymity set size,
the CDF plots would exhibit a single vertical line.
In contrast, the simulation results show that there are many streams with very little anonymity remaining,
just like there are many streams with rather strong anonymity remaining and many streams in between.
While a result like this was to be expected due to the fact that
Tor has no means of \emph{ensuring} a certain level of anonymity,
it still shows a surprisingly clear weakness of the system
because users cannot reasonably expect a certain level of anonymity,
as far as the stream-based anonymity degradation over time is concerned.

Having seen in Section~\ref{sec:synthetic-streams}
that a stream's anonymity is heavily influenced by its length,
we next investigate the relationship between stream lengths and anonymity in the simulated networks.
Figure~\ref{subfig:orig} shows CDF plots of the anonymity sets,
subdivided into quantiles based on their length (number of cells),
from percentiles~0--10~(black), to~90--100~(orange).
The resulting anonymity distributions clearly point out that
stream length is again a decisive factor for anonymity.
Long-lasting streams are massively disadvantaged
and generally achieve only little anonymity.
In contrast, short streams are much harder to de-anonymize
and have anonymity sets that are multiple times larger.
These results again support our observations from Section~\ref{sec:synthetic-streams}:
Long streams suffer from reduced anonymity but act as cover traffic
providing stronger anonymity for shorter streams.

\paragraph{Anonymity in Scaled Networks:}
We leverage our simulation and analysis workflow
to evaluate not only the Tor network as is today
but also variations of the network
that could result from future network growth with different characteristics.
Evaluating scaled Tor networks allows us to investigate
the influence of network topology on anonymity.
Understanding the trade-offs of different topological properties
may provide important insights on how to shape and prepare future network evolution.
In addition to a snapshot of today's Tor network,
we therefore carried out simulations of the Tor network being scaled by +100\% vertically
as well as by +100\% horizontally.
By \enquote{vertical} growth we mean that the bandwidth capacity of each relay is increased accordingly.
Likewise, \enquote{horizontal} growth is achieved by adding new, additional relays to the existing network.
The scaling methodology of \emph{torsynth} is detailed in~\cite{DBLP:conf/IEEEares/DopmannT23}.
For simulations, we also grow the number of active users on the network by +100\%
so the networks differ only in their topology, and not in the amount of available resources per user.

We first consider the case of horizontal growth.
Table~\ref{tab:case-study-metrics} and Figure~\ref{fig:case-study-cdf}
show that increasing the number of relays has a strong impact on anonymity as defined by progressive pruning.
In fact, with double the number of relays, the average anonymity size is also exactly doubled.
This trend is intuitively reasonable because doubling the number of relays
also implies doubling the number of available exits
which are the potential anonymity set candidates.
However, it is interesting to see that this increase of average anonymity set size
appears to be exactly proportional to the number of exit relays.
This is somewhat surprising because, essentially,
the number of exits only directly defines the \emph{maximum} anonymity set size
which wouldn't necessarily have to correspond linearly to the average anonymity set size.
Even if the anonymity set sizes grow with horizontal scaling of the network,
it has to be noted that the standard deviation grows massively, too.
In our experiments, it also exactly doubles.
We regard this as a problematic property
because it means that anonymity is distributed much more unevenly among users of the network.

Looking at the vertically upscaled network instead paints a rather different picture.
Due to the fact that the number of exits---and thus anonymity set candidates---has not increased,
the average anonymity set size has not doubled as in the case of horizontal growth.
While this was to be expected, the vertical growth (and the increased load)
of the network still results in a clear increase of anonymity by approximately 30\%.
Combined with our previous observations,
this makes clear that there is no single parameter that determines the anonymity level.
Instead, we see that there are two primary factors for anonymity as defined by our metric:
topology on the one hand (determining the maximum number of potential anonymity set members)
and traffic load on the other.
Also note that, compared to the original Tor network,
the vertically scaled instance exhibits \emph{less} variation in anonymity set size among the streams
which is in contrast to the previously observed increase in the case of horizontal growth,
and which might be a desirable system property.

All in all, we can conclude from the simulations of the Tor network which we carried out
that there are two fundamental factors for anonymity in the network:
Firstly, an increased number of relays also increases the number of (topological) candidates
for the anonymity set.
And secondly, the mere increase in network load
also improves anonymity, albeit much less effectively.

\section{Conclusion} \label{sec:conclusion}

In this work, we analyzed the inherent vulnerability
of stream-based anonymous communication to traffic analysis.
We specifically focused on intersection attacks,
which exploit fundamental weaknesses of streams
compared to independent messages.
To this end, we introduced \emph{progressive pruning},
a methodology to quantify stream anonymity in the face of such attacks,
enabling a differentiated assessment beyond a binary view of susceptibility.
This approach facilitates comparing anonymity across system variants,
aiding the design of new ACNs that account for attackers' costs.

We conducted simulations to examine the impact of stream properties
in various scenarios.
Our findings highlight that stream length relative to others
significantly influences susceptibility:
longer streams are more easily distinguished,
reducing their anonymity,
but also serve as cover traffic for shorter streams.

Additionally, we demonstrated the application of progressive pruning
to large real-world networks,
By using Tor as a case study, we observed that horizontal scaling
improves anonymity by increasing anonymity set sizes
under a global adversary model, but anonymity is distributed very unevenly.
On the other hand, when considering a vertically scaled network,
we can see that an increased number of users is beneficial for anonymity.

Despite the inherent vulnerability to intersection attacks,
our findings provide a methodology to systematically explore mitigation strategies
increasing the difficulty faced by adversaries.

\printbibliography

\end{document}